\documentclass[sigconf]{acmart}

\pdfoutput=1

\usepackage{amsmath,amsfonts,bm}

\def\eqref#1{equation~\ref{#1}}

\def\1{\bm{1}}

\def\vh{{\bm{h}}}

\def\vo{{\bm{o}}}
\def\vp{{\bm{p}}}

\def\vs{{\bm{s}}}

\def\vw{{\bm{w}}}
\def\vx{{\bm{x}}}

\def\vz{{\bm{z}}}

\def\mA{{\bm{A}}}

\def\mH{{\bm{H}}}

\def\mO{{\bm{O}}}

\def\mS{{\bm{S}}}

\def\mW{{\bm{W}}}
\def\mX{{\bm{X}}}

\def\mZ{{\bm{Z}}}

\DeclareMathAlphabet{\mathsfit}{\encodingdefault}{\sfdefault}{m}{sl}
\SetMathAlphabet{\mathsfit}{bold}{\encodingdefault}{\sfdefault}{bx}{n}

\newcommand{\R}{\mathbb{R}}

\newcommand{\relu}{\mathrm{ReLU}}

\DeclareMathOperator*{\argmin}{arg\,min}

\usepackage{booktabs} %
\usepackage[colorinlistoftodos]{todonotes}
\usepackage{algorithmic}
\usepackage[linesnumbered, ruled, vlined]{algorithm2e}
\usepackage{multirow}
\usepackage{tablefootnote}
\usepackage{graphicx}
\usepackage{subfigure}
\usepackage{flushend}
\usepackage{color}
\usepackage{enumitem}
\usepackage{array, makecell}
\usepackage{footmisc} 
\usepackage{utfsym} 

\usepackage{xcolor}
\usepackage{bbm}

\newcommand{\hide}[1]{} %
\newcommand{\vpara}[1]{\vspace{0.07in}\noindent\textbf{#1 }}

\newtheorem{definition}{Definition}
\newtheorem{problem}{Problem}

\newcommand{\Real}{\ensuremath{\mathbb{R}}}  %
\newcommand{\beq}[1]{{\begin{equation}#1\end{equation}}}
\newcommand{\beqn}[1]{{\begin{eqnarray}#1\end{eqnarray}}}

\newcommand{\model}{P2TAG\xspace}
\newcommand{\modelskip}{P2TAG}

\usepackage{flushend}
\usepackage[normalem]{ulem}

\AtBeginDocument{%
  }

\setcopyright{rightsretained}
\copyrightyear{2024}
\acmYear{2024}
\acmDOI{10.1145/3637528.3671952}

\acmConference[KDD '24]{Proceedings of the 30th ACM SIGKDD Conference on Knowledge Discovery and Data Mining}{August 25--29, 2024}{Barcelona, Spain}
\acmBooktitle{Proceedings of the 30th ACM SIGKDD Conference on Knowledge Discovery and Data Mining (KDD '24), August 25--29, 2024, Barcelona, Spain}

\acmISBN{979-8-4007-0490-1/24/08}

\begin{document}

\title{Pre-Training and Prompting for Few-Shot Node Classification on Text-Attributed Graphs}

\author{Huanjing Zhao$\dagger$}
\affiliation{
\institution{Tsinghua University}
\city{Beijing}
\country{China}
}
\email{hwangyeong@mail.tsinghua.edu.cn}

\author{Beining Yang$\dagger\ddagger$}
\affiliation{
\institution{University of Edinburgh}
\city{Edinburgh}
\country{UK}
}
\email{B.Yang-32@sms.ed.ac.uk}

\thanks{$\dagger$~Both authors contributed equally to this research.}
\thanks{$\ddagger$~This work was done when the author was interned at Zhipu AI.}

\author{Yukuo Cen}
\affiliation{
\institution{Zhipu AI}
\city{Beijing}
\country{China}
}
\email{yukuo.cen@zhipuai.cn}

\author{Junyu Ren}
\affiliation{
\institution{Tsinghua University}
\city{Beijing}
\country{China}
}
\email{renjy19@mails.tsinghua.edu.cn}

\author{Chenhui Zhang}
\affiliation{
\institution{Zhipu AI}
\city{Beijing}
\country{China}
}
\email{chenhui.zhang@zhipuai.cn}

\author{Yuxiao Dong}
\affiliation{
\institution{Tsinghua University}
\city{Beijing}
\country{China}
}
\email{yuxiaod@tsinghua.edu.cn}

\author{Evgeny Kharlamov}
\affiliation{
\institution{Bosch Center for Artificial Intelligence}
\city{Renningen}
\country{Germany}
}
\email{evgeny.kharlamov@de.bosch.com}

\author{Shu Zhao}
\affiliation{
\institution{Anhui University}
\city{Anhui}
\country{China}
}
\email{zhaoshuzs@ahu.edu.cn}

\author{Jie Tang$^*$}
\affiliation{
\institution{Tsinghua University}
\city{Beijing}
\country{China}
}
\email{jietang@tsinghua.edu.cn}

\thanks{$^*$~Corresponding author: JT}

\thanks{$^1$~Our code is available at \url{https://github.com/THUDM/P2TAG}}

\renewcommand{\authors}{Huanjing Zhao, Beining Yang, Yukuo Cen, Junyu Ren, Chenhui Zhang, Yuxiao Dong, Evgeny Kharlamov, Shu Zhao, Jie Tang}
\renewcommand{\shortauthors}{Huanjing Zhao et al.}

\begin{abstract}
    The text-attributed graph (TAG) is one kind of important real-world graph-structured data with each node associated with raw texts. 
    For TAGs, traditional few-shot node classification methods directly conduct training on the pre-processed node features and do not consider the raw texts. 
    The performance is highly dependent on the choice of the feature pre-processing method.
    In this paper, we propose \modelskip$^1$, a framework designed for few-shot node classification on TAGs with graph pre-training and prompting.
    \model first pre-trains the language model (LM) and graph neural network (GNN) on TAGs with self-supervised loss. 
    To fully utilize the ability of language models, we adapt the masked language modeling objective for our framework. 
    The pre-trained model is then used for the few-shot node classification with a mixed prompt method, which simultaneously considers both text and graph information. 
    We conduct experiments on six real-world TAGs, including paper citation networks and product co-purchasing networks. 
    Experimental results demonstrate that our proposed framework outperforms existing graph few-shot learning methods on these datasets with +18.98\% $\sim$ +35.98\% improvements.

\end{abstract}



\maketitle

\section{Introduction}

The few-shot node classification task involves identifying the classes of nodes in a given graph structure using only a limited number of labeled examples. This task has practical applications in areas such as social network analysis, recommendation systems, and more. Inspired by the successful experiences in the field of Computer Vision (CV), several works, such as G-Meta and TENT~\cite{yao2020graph, ding2020graph, huang2020graph, wang2022task}, apply meta-learning to graphs to address the few-shot node classification problem. These methods learn transferable knowledge from meta-tasks, enabling rapid adaptation to new, unseen labels. Unlike images, graphs represent a form of structured data. Through graph pre-training, models can also accumulate a substantial amount of domain-specific knowledge.

Recently, a series of graph pre-training methods~\cite{thakoor2022large, zhu2020deep, zhang2021canonical, zheng2022rethinking, yang2023does} emerged with self-supervised learning (SSL) to yield generalized node representations in the absence of labels.
These methods mainly include contrastive and generative ones. 
Contrastive SSL methods  utilize data augmentation to generate multiple views of data for contrastive loss. 
Generative SSL methods such as GraphMAE~\cite{HouLCDYW022} aim to reconstruct the (masked) attributes of the graph. 
These self-supervised learning methods have greatly contributed to many aspects, such as graph augmentation and graph-based pretext tasks. 
They only consider part of the self-supervised learning on TAGs, making the training of GNNs independent of LMs text encoding.

The nodes in the graph usually contain rich textual information \cite{ji2021survey} such as titles and abstracts in citation networks. 
Some recent works \cite{ChienCHYZMD22, zhao2022learning, yang2021graphformers} attempt to integrate the training of LMs and GNNs. 
These methods directly deal with the original texts and the topological structure in the graph, 
and achieve better performance on the downstream tasks. 
GIANT \cite{ChienCHYZMD22} aims to obtain stronger textual representations by the language model with neighborhood prediction in TAGs. 
Although GIANT utilizes graph information to fine-tune the language models, the text encoding and graph propagation steps are still independent. 
GraphFormer~\cite{yang2021graphformers} proposes a nested architecture of LMs and GNNs and jointly trains them with a link-prediction-based objective. 
GLEM~\cite{zhao2022learning} alternatively optimizes LMs and GNNs using label information from the downstream tasks. 

In few-shot node classification, prompting helps guide the pre-trained model to generalize from a few examples by providing specific input cues. 
Prog~\cite{sun2023all} constructs prompts at the graph level, which is not directly applicable to TAGs. 
With the leapfrog development of large language models (LLMs), some works attempt to construct prompts by processing raw texts in the TAG, the name we refer to as hard text prompt, exemplified by TAPE~\cite{he2023harnessing} and ENG~\cite{yu2023empower}. 
GPrompt \cite{xuanwen2023prompt} leverages the GNN model as a downstream adapter, using hard text prompts to improve adaptation efficiency.
These works rely on the quality of constructed prompts.
It is highly challenging since there are both text (LM side) and structure information (GNN side) on TAGs.
G2P2~\cite{wen2023augmenting} tries to solve this problem solely from the LM side, named soft text prompt. They reweight the pre-trained GNN embedding by the neighboring nodes' texts and label texts. However, the potential capability of GNN is neglected in this way and may lead to sub-optimal results.

\begin{table}[t]
\centering
\captionsetup{skip=6pt}
\caption{Related works of few-shot learning on TAGs. }
\label{tab:related_work}
\begin{tabular}{llll} 
\toprule
Method     & Few-shot & Text & Prompt Type          \\
\midrule
G-Meta~\cite{huang2020graph}     & \usym{2714}      & -    & -                \\
TENT~\cite{wang2022task}       & \usym{2714}      & -    & -                \\
Prog~\cite{sun2023all} & \usym{2714}      & -    & Graph Prompt     \\
TAPE~\cite{he2023harnessing}       & -        & \usym{2714}  & Hard Text Prompt      \\
ENG~\cite{yu2023empower}        & \usym{2714}      & \usym{2714}  & Hard Text Prompt      \\
GPrompt~\cite{xuanwen2023prompt}         & \usym{2714}      & \usym{2714}  & Hard Text Prompt      \\
G2P2~\cite{wen2023augmenting}       & \usym{2714}      & \usym{2714}  & Soft Text Prompt  \\
\midrule
\model \textbf{(Ours)}       & \usym{2714}      & \usym{2714}  & Mixed Prompt     \\
\bottomrule
\end{tabular}
\vspace*{-12pt}
\end{table}

\vpara{Present work: \model. }
In this paper, we propose the \textbf{P}re-training and \textbf{P}rompting for few-shot node classification on \textbf{TAG}s.
In the pre-training phase, different from previous works, our framework integrates the LM and GNN effectively with joint training in a self-supervised way. 
Some solutions are proposed to address the challenges of integration of LMs and GNNs. 
We incorporate the encoder of GNNs as a supplement to LMs and keep the original self-supervised loss of the LM model, i.e., the masked language modeling objective, 
which is a very hard pretext task and can alleviate the over-fitting issue. 
To train the GNNs and the LMs jointly, our framework first samples mini-batch subgraphs through a random-walk-based sampler. 
Each mini-batch will be fed to the LMs for text encoding, and the output of the classification token (i.e., [CLS] token) will be considered as the representation of the node (or the text sequence). 
The output embeddings of [CLS] tokens will be fed to the GNNs to aggregate the information from neighbors through graph propagation. 
For the masked language modeling, 
we incorporate the aggregation of neighbor information into the outputs of [MASK] tokens. 
Finally, the cross-entropy loss of the outputs and the original tokens is computed. 

In the prompting phase, to bridge the gap between pre-train tasks and the downstream ones, we process with a mixed approach of LM and GNN, which differs from previous works as shown in Table~\ref{tab:related_work}. Specifically, we jointly concatenate the \textit{label text initialized graph prompt} and a simple \textit{LM output initialized text prompt} to mimic the training paradigm. As illustrated in Figure~\ref{fig:motivation_label_as_init}, by leveraging the label text embedding, we can easily align the node to the downstream label space without 
discrete and intuitive hand-craft prompts. The empirical results presented in Section~\ref{sec:effct_of_prompts} attest to the superior efficacy of our prompt design, which achieves a substantial improvement of 11.1\% $\sim$ 25.2\% over the existing soft text prompted G2P2 method~\cite{wen2023augmenting}.

The key contributions of our framework are summarized in the following. 

\begin{itemize} [leftmargin=15pt]
    \item We propose a unified framework to jointly train language models and graph neural networks via the masked language modeling objective. 
    \item We propose a new graph-text mixed prompt learning method with text and structure information to mitigate the gap between pre-training and few-shot downstream tasks. 
    \item Our proposed \model achieves an improvement with +18.98\% $\sim$ +35.98\% over meta-learning methods across six TAG datasets.
\end{itemize}

\section{Preliminaries}
\label{sec:background}
In this section, we introduce the background of our paper including text-attributed graph and few-shot node classification.

\vpara{Notations.}
Denote a graph $G=(\mathcal{V},\mathcal{E})$, where $\mathcal{V}$ is a set of $n$ nodes and $\mathcal{E}$ is a set of edges between nodes. 
$\mA \in \R^{n\times n}$ is the adjacency matrix where its entry $\mA(i, j) \geq 0$,
if nonzero, denotes there is an edge between node $i$ and $j$ with edge weight $\mA(i, j)$. 
In practice, the network could be either directed or undirected. If $G$ is directed, we have $\mA(i, j) \nequiv \mA(j, i)$; if $G$ is undirected, we have $\mA(i, j)\equiv \mA(j, i)$.

\subsection{Text-Attributed Graph}

\begin{definition}[Text-Attributed Graph]
A \textbf{text-attributed graph (TAG)} is a graph $G=(\mathcal{V},\mathcal{E}, \mathcal{S})$, where each node $v_i\in \mathcal{V}$ is associated with a text sequence $\vs_i \in \mathcal{S}$ and $\mathcal{E}$ represents the set of edges between nodes. 
\end{definition}

Given the above definition, we can formally define our problem for self-supervised learning on TAG.

\begin{problem}[Self-supervised Learning on TAGs]
Given a TAG $G=(\mathcal{V},\mathcal{E},\mathcal{S})$, the problem of \textbf{self-supervised learning (SSL) on TAG} is to learn a unified low-dimensional space representation of each text-attributed node $v_i\in V$ with text sequence $\vs_i$. The goal is to find a function $f: \mathcal{V}\to \mathbb{R}^d$, where $d\ll |\mathcal{V}|$.
\end{problem}

Previous methods implement self-supervised learning on TAG through separated text encoders and GNNs. Built upon these works, we explore self-supervised learning by integrating LMs and GNNs. 
The joint training for the problem of SSL on TAGs can be formulated as follows:
\beq{
    \nonumber
    \theta_1^*, \theta_2^*, \theta_3^* = \argmin_{\theta_1, \theta_2, \theta_3} \mathcal{L}_{SSL} \left(f_{\theta_1}^{GNN}, f_{\theta_2}^{LM}, f_{\theta_3}^{Head}, G \right), 
}
where $f_{\theta_1}^{GNN}$, $f_{\theta_2}^{LM}$, and $f_{\theta_3}^{Head}$ are the GNN encoder, LM encoder, and the prediction head for the SSL. 
Then we can obtain the final embeddings via the following:
\beq{
    \nonumber
    \mZ = f_{\theta_1^*}^{GNN}\left(G, \mX \right), \text{ where } \mX = f_{\theta_2^*}^{LM} \left(\mathcal{S} \right).
}

\subsection{Few-shot Node Classification}
Few-shot node classification are tasks that involve training a machine learning model with a limited number of labeled nodes, and subsequently predicting the classes of other nodes based on these initial labels. The $N$-way $K$-shot few-shot node classification tasks $\mathcal{T}$ on a graph is described as follows with the node label set $\mathcal{C}$. Each task $\mathcal{T}_t \in \mathcal{T}$ selects $N$ labels from $\mathcal{C}$, forming a subset $\mathcal{C}_t = \{ \mathcal{C}_1, \mathcal{C}_2,..., \mathcal{C}_N \} \in \mathcal{C}$. The $\mathcal{T}_t$ consists of two parts as follows:
\beq{
    \nonumber
    \mathcal{T}_t = \{ \mathcal{S}_t, \mathcal{Q}_t \},
}where $\mathcal{S}_t$ means the support set and $\mathcal{Q}_t$ means the query set. The machine learning model is trained on $\mathcal{S}_t$ and subsequently evaluated on $\mathcal{Q}_t$. Both the support set and the query set consist of nodes and their corresponding labels, which shown as follows:
\beqn{
    \nonumber
    \mathcal{S}_t = \{ (v_1, c_1), (v_2, c_2), ..., (v_{N \times K}, c_{N \times K}) \}, \\
    \nonumber
    \mathcal{Q}_t = \{ (v'_1, c'_1), (v'_2, c'_2), ..., (v'_{N \times Q}, c'_{N \times Q}) \},
}where $v$ and $v'$ represent nodes, their corresponding labels are denoted as $c$ and $c'$. In the support and query sets, each label $\mathcal{C}_i \in \mathcal{C}_t$ associated with $K$ nodes and $Q$ nodes, respectively. For the support set, there exists $c_{(i-1)*K + j} \in \mathcal{C}_i$, where $i \in \{1,2,..,N\}$ and $j \in \{1,2,..,K\}$. Similarly, there exists $c'_{(i-1)*Q + j} \in \mathcal{C}_i$, where $i \in \{1,2,..,N\}$ and $j \in \{1,2,..,Q\}$ for the query set.

\section{Methodology}
\label{sec:method}
\begin{figure*}
    \centering
    \includegraphics[width=0.9\textwidth]{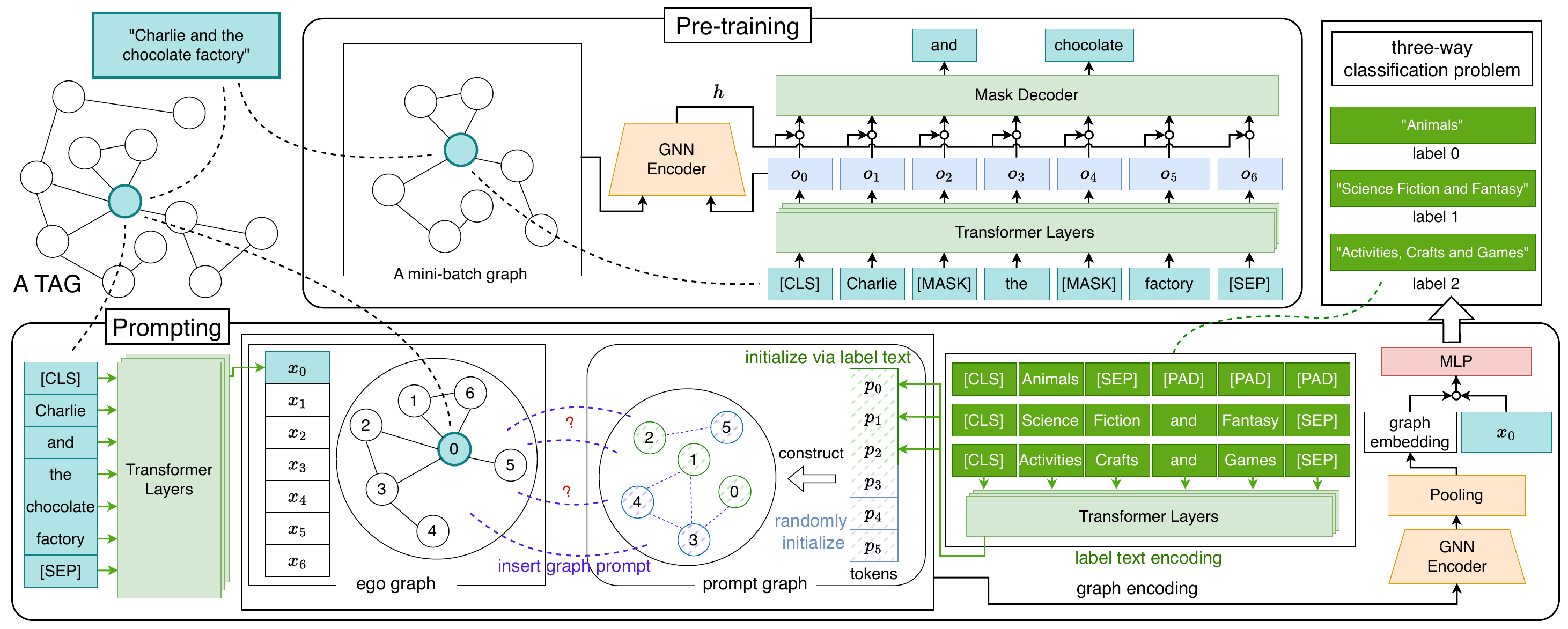}
    \caption{Our proposed framework \model. \textmd{A toy example illustrating the 3-way classification of children's books. (1) In the pre-training phase, we jointly train the LM and GNN using a masked language modeling objective. For the GNN module, a subgraph-based sampler employing random walks generates mini-batch subgraphs for training. (2) In the prompting phase, we construct the ego graph for each target node and generate node features with the pre-trained LM. Graph tokens are utilized to learn the graph structure and create a prompt graph. These tokens are initialized either by encoding label text or through a random initialization process.}}
    \label{fig:framework}
\vspace*{-5pt}
\end{figure*}

We tackle few-shot node classification by adhering to the pre-training and prompting paradigm, as depicted in Figure~\ref{fig:framework}. 
To better leverage the unique characteristics of the TAGs, we propose a new self-supervised framework for pre-training. 

\subsection{Pre-training Framework}

In this part, we introduce our proposed pre-training framework in detail. 
For self-supervised learning on TAGs, previous works usually take two separate steps. 
The first step is to encode the raw texts to node features using bag-of-words, word2vec~\cite{mikolov2013efficient}, or pre-trained language models~\cite{devlin2018bert, lan2020albert, yang2019xlnet, clark2020electra, liu2019roberta, du2022glm, he2021deberta}. %
Most graph self-supervised methods~\cite{thakoor2022large, zhu2020deep, zhang2021canonical, zheng2022rethinking, HouLCDYW022} use the node features pre-processed in the first step to construct a self-supervised objective. 
The two-step process can bring non-negligible information loss. 
To address this issue, we propose an end-to-end self-supervised framework to train directly on the TAGs. 
Inspired by the recent prosperity of pre-trained language models (LM), we choose an effective LM to encode the raw texts.
To model the relations between nodes (texts), we can utilize powerful graph neural networks (GNNs). 
As illustrated in Figure~\ref{fig:framework}, our framework consists of two core modules, including a pre-trained language model and a GNN encoder. 
The GNN encoder is to help make better node representations. 
In our framework, we mainly use the DeBERTa-base~\cite{he2021deberta} with 100M parameters as the LM of our framework. 
DeBERTa utilizes two techniques based on BERT~\cite{devlin2018bert} and RoBERTa~\cite{liu2019roberta} and significantly improves the performance on the natural language understanding and generation tasks. 
The choice of the LMs is flexible, and we also explore other LMs in our experiments. 

\textit{The pre-training of LMs and GNNs is much more challenging because we need to 1) choose an appropriate self-supervised training objective to avoid over-fitting and 2) sample small mini-batches to address the high computational and space costs of large LMs. }

\vpara{Self-supervised training objective.}
The training objective plays an important role in self-supervised learning. 
Different from conventional graph self-supervised learning, the design of self-supervised objective for TAGs are more challenging.
Our model architecture contains two different modules with different scales of model parameters (large LM v.s. small GNN), making the training much more difficult. 
Simple self-supervised objectives will probably make the model overfitted. 
To make a harder self-supervised objective, we adopt the masked language modeling (MLM) introduced in~\cite{devlin2018bert} as our objective. 
Given a node $v_i$, we first feed the text sequence $\vs_i = [T_1, T_{2}, \dots, T_{n_i}]$ into a language model to get the hidden representations associated with $v_i$ where $T_t$ is $t$-th token of $\vs_i$ and $n_i$ is the length of $\vs_i$.
In the training stage, we randomly mask a pre-defined portion of tokens by replacing them with a special token named [MASK].
We use $\vs'_i = [T'_{1}, T'_{2}, \dots, T'_{n_i}]$ to denote the text sequence after masking and each token $T'_{t}$ is a random variable with the following distribution:
\begin{equation}
  \Pr \left(T'_{t} = [MASK]\right) = p \text{ and } \Pr \left(T'_{t} = T_{t}\right) = 1 - p.
\end{equation}
Here, $p \in (0, 1)$ is a hyper-parameter representing the mask rate.
When working with BERT-like language models, we usually add a starting token (e.g., [CLS]) and an ending token (e.g., [SEP]) to the sequence.
Therefore, each node is represented as a sequence of hidden vectors:
\begin{equation}
  [\vo_{0}, \vo_{1}, \dots, \vo_{n_i},\vo_{n_i + 1}] = f^{LM} \left([[CLS], T'_{1}, T'_{2}, \dots, T'_{n_i}, [SEP]]\right),
\end{equation}
where $\vo_{t} \in \Real^d$ is the hidden representations of $t$-th token of $\vs'_i$.
Notice that the first hidden vector $\vo_0$ corresponds to the special token [CLS] and can be treated as the ``summary'' representation of the whole text sequence $\vs_i$.
To capture the correlations among nodes, we then use a graph neural network to propagate the hidden representations of nodes where the input of node $v_i$, denoted as $\vx_i$, is the first hidden vector $\vo_0$ of $\vs_i$.
The node representations after passing a GNN encoder are denoted as $\mH = f^{GNN} (G, \mX)$.
The propagated node presentations are exploited to construct the self-supervised training objective.

For each node $v_i$, we concatenate the hidden representation $\vh_i$ with the output vector $\vo_t$ of each token and then feed the concatenated vector to an MLP as: 
\begin{equation}\label{eq:concat_feature}
    \vz_t = \operatorname{MLP}([\vh_i^\top, \vo_t^\top]), \text{where}\ t=0,\cdots ,(n_i+1).
\end{equation}

The objective of the masked language model (MLM) is to predict the masked tokens in the original text sequence $\vs_i$, which can be expressed as:
\begin{equation}
  \label{eq:mlm_loss}
  \mathcal{L}_i = \sum_{t=1}^{n_i} \mathbbm{1} \left(T'_{t} = [MASK]\right) \log \Pr \left(T_{t} \mid \vz_{t}\right).
\end{equation}

The indicator $\mathbbm{1} (T'_{t} = [MASK])$ ensures that the loss is only applied on masked tokens.
We can control the difficulty of the SSL task via a flexible hyperparameter, i.e., mask rate.

\begin{algorithm}[thbp]
    \caption{The pre-training stage of our framework. }
    \label{alg:training}
    \textbf{Input} TAG $G=(\mathcal{V}, \mathcal{E}, \mathcal{S})$, training steps $M$, mask rate $p$, GraphSAINT sampler $\operatorname{Sample}$\;
    \textbf{Output} Model parameters $\theta$\ of both the LM and GNN;

    Pre-processing: compute normalization coefficient $\alpha$ and $\lambda$\ according to GraphSAINT sampler~\cite{ZengZSKP20}\;
    \For {step $\leftarrow$ 1 to $M$} {
        $G_s = (\mathcal{V}_s, \mathcal{E}_s) \gets \operatorname{Sample} (G)$\;
        $\mS \gets \{\vs_i: v_i \in \mathcal{V}_s\}$\;
        $\mS' \gets \{ \operatorname{MaskToken} (\vs_i, p): \vs_i \in \mS \}$\;
        $\mO \gets f^{LM} (\mS')$\;
        $\mH \gets f^{GNN} (G_s, \mO[:, 0])$\; %
        $\mO \gets \operatorname{Concat}(\mO, \mH)$\;
        $\mZ \gets \operatorname{MLP}(\mO)$\;
        Compute the loss via Equation~\ref{eq:mlm_loss}\; %
        Update $\theta$ based on the gradient of the loss\;
    }
    \Return{Model parameters $\theta$}\;
    \normalsize
\end{algorithm}

\subsection{Mini-batch Training}\label{sec:minibatch}
Due to the heavy language models, we need to adopt a mini-batch training strategy even if we train our model on small graph datasets. 
There are several genres of mini-batch training techniques for GNNs. 
To trade off the efficiency and flexibility, we choose a subgraph-based method, GraphSAINT~\cite{ZengZSKP20}, as our training strategy. 
At every training step, GraphSAINT constructs a mini-batch by sampling a subgraph from the original graph and generates node representations according to the sampled subgraph.
In this work, we adopt the random walk sampler to preserve the connectivity of the whole graph.
The sampling process starts by selecting $r$ root nodes from the whole node set $\mathcal{V}$ uniformly at random.
Starting from each root node, a random walk of length $l$ is sampled from the original graph structure.
We then have a sampled node set $\mathcal{V}_s$ by adding all the nodes that occurred in the random walks and the subgraph $G_s$ induced by $\mathcal{V}_s$ is used in the GNN encoder to generate node representations in the minibatch.

\subsection{Graph-Text Mixed Prompt Learning}\label{sec:graph_prompt}
To prevent tuning the whole pre-trained model in the few-shot downstream tasks, we adopt the prompt learning paradigm, \textit{i.e.}, using a few trainable parameters to be the substitution of full-scale model parameters. 
\textbf{The goal of prompt design is to mitigate the gap between the pre-train objective and the downstream tasks}. It is highly challenging since there are both text (LM side) and structure information (GNN side) on TAG graphs. 
We try to investigate the joint prompting method, namely the \textit{graph prompt} and the \textit{text prompt}, the detailed design as follows.
To simplify the discussion, we start with a target node $v$ (\textit{e.g.}, the $0$-th node in Figure~\ref{fig:framework}). We further introduce the concept of the ego graph node set $\mathcal{V}_{ego}$. We select up to 100 first-order neighbors of the given node, along with the node itself, to form the node set of the ego graph. Then we define the induced ego graph from $\mathcal{V}_{ego}$ as $G_{ego}$.

\subsubsection{Graph prompt design}
Inspired by the recently proposed graph prompt literature~\cite{sun2022gppt, sun2023all}, we aim to bridge gap through a small yet informative prompt graph. 
We first craft a prompt graph $G_{p} = (\mathcal{V}_{p}, \mathcal{E}_{p})$ serving as an ancillary component connected to the target ego node set $\mathcal{V}_{ego}$. $G_{p}$ comprises $\|G_{p}\|$ nodes (\textit{i.e.}, tokens), the internal edges between tokens can be defined as:
\begin{equation}\label{eq:inner_structure}
\mathbf{A}_{G_p}(i, j) =
\begin{cases}
1 & \text{if } \operatorname{Sim}(\vp_i, \vp_j) \ge \sigma_{inner}, \\
0  & \text{otherwise},
\end{cases}
\end{equation}
where the $\operatorname{Sim}(\cdot)$ denotes the similarity function, and the $\vp_i$ denotes the feature of $i$-th tokens; in this paper, we use dot product as the implicit operation. Then, let the $\vx_k$ denotes on the $k$-th node feature of the $G_{ego}$, the inter-connectivity between $G_{p}$ and $G_{ego}$ is given by:
\begin{equation}\label{eq:outer_structure}
\mathbf{A}_{G_{ego}, G_{p}}(k, j) =
\begin{cases}
1 & \text{if } \operatorname{Sim}(\vx_k, \vp_j) \ge \sigma_{inter}, \\
0  & \text{otherwise}.
\end{cases}
\end{equation}
Here, the $\sigma_{inner}$ and $\sigma_{inter}$ are two hyper-parameters indicating the pruning value, \textit{i.e.}, the entries on the matrix higher than $\sigma_{inner}$ and $\sigma_{outer}$ are formed as edge respectively.

\begin{figure}[H]
    \centering
    \includegraphics[width=0.47\textwidth]{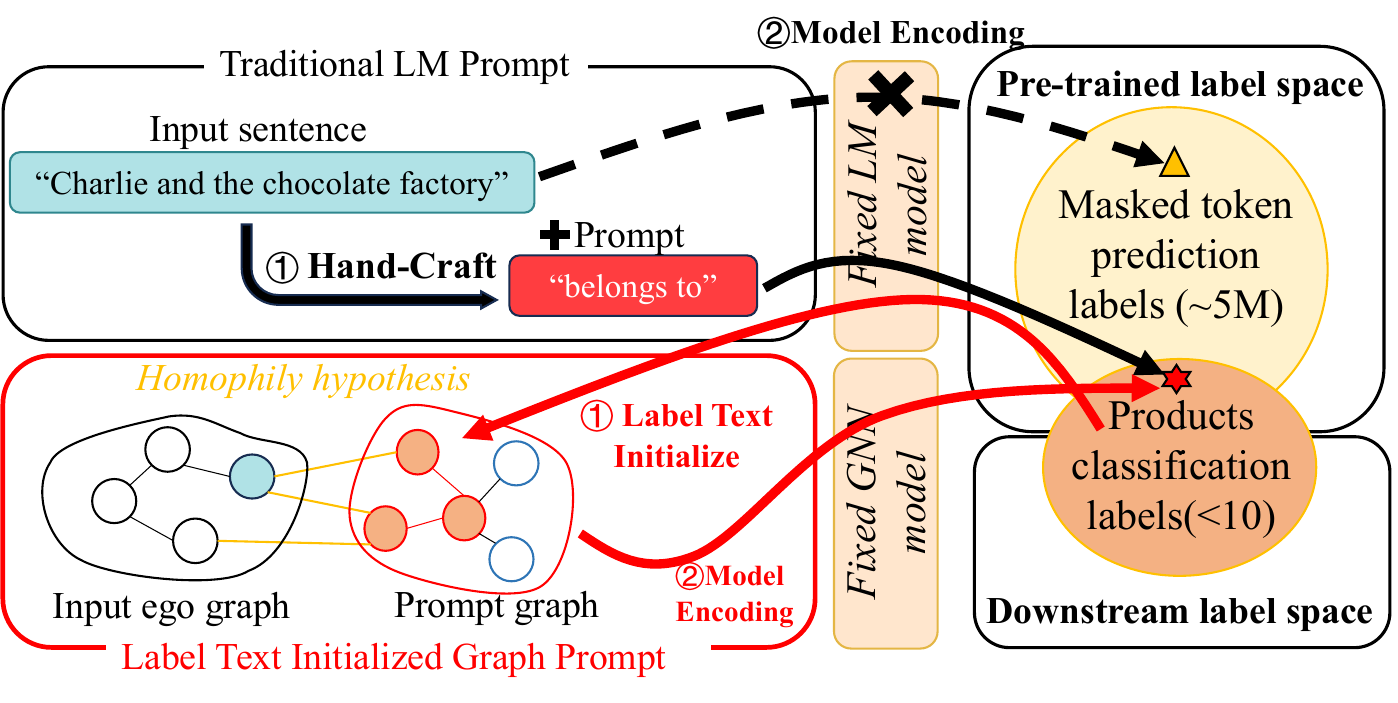}
    \caption{Initialize tokens via the label text. \textmd{Crafting the prompt graph via downstream label texts under the} homophily hypothesis, \textmd{thereby making the target node embedding more aligned with the downstream label space.}}
    \label{fig:motivation_label_as_init}
    \vspace*{-10pt}
\end{figure}

\vpara{Initialize tokens via the label text.} While the prompt graph servers as an efficient substitute for full-scale model parameters, we empirically find that the naive $G_{p}$ performs sub-optimally. We attribute this to the inconsistency in the pre-trained task objectives between our masked token prediction task and previous graph-based ones \cite{sun2022gppt, sun2023all} (\textit{e.g.}, the link prediction task).

Referencing Figure~\ref{fig:motivation_label_as_init} as an example, during the pre-training phase where token prediction tasks are undertaken, the model is equipped to align input with the pre-trained label space. Consequently, the prompt should slightly modify the input distribution to approximate the downstream label space more closely. Hence, unlike traditional cumbersome and  hand-crafted text prompt templates, we discovered that employing label text embeddings for prompt graph initialization is more direct and efficient. \textbf{Under the homophily hypothesis, the target node's embedding is immediately positioned more closer to the desired label space.}

Specifically, we utilize the pre-trained $f^{LM}$ as our label text encoder, each label text (\textit{e.g.}, ``Animals'', ``Science Fiction and Fantasy'', and ''Activities, Crafts and Games'') can be represented as $s_{y_i}$. Then we can formalize the initialization of $G_{p}$ as follows:

\begin{equation}
    \vp_i^{(0)} = 
    \begin{cases}
    f^{LM}(\vs_{y_i}) & \text{if } i \le N, \\
    \text{Random Initialization} & \text{otherwise}.
    \end{cases}
\end{equation}
Here, the $\vp_i^{(0)}$ denotes the initial embedding of i-th prompt node, and for the additional prompt nodes, we adopt the random initilization. Then the inner structure of each nodes will be constructed through the Eq.~\ref{eq:inner_structure}. The links among the prompt nodes and the target ego graph $G_{ego}$ are built by the Eq.~\ref{eq:outer_structure}.

\subsubsection{Text prompt design}

As shown in Equation~\ref{eq:concat_feature}, we use the joint representation of the LM model output (text information) and the GNN model output (structure information). Therefore, to align the pre-train and downstream tasks, we use a similar strategy to concat both output features in the downstream scenario.

Then, instead of making intricate text-based prompts, we use a rather simple way: letting the second concat feature be a trainable parameter, namely $\vw_{t}\in \mathbb{R}^{1\times d}$, \textit{i.e.}, treating the LM model output itself as a tunable parameter. The initialization process can be defined similarly as:
\begin{equation}
    \vw_{t}^{0} = f^{LM} (\vs_{v}),
\end{equation}
the $\vs_v$ shows the text sequence with the target node $v$.
Our ablation study in Section~\ref{sec:effct_of_prompts} shows our text prompt's effectiveness. 

\subsubsection{Final prompt learning forward function}Assuming we have a node ego graph $G_{v}$ and its text $\vs_v$, the prompt learning forward function is:
\begin{equation}
\tilde{\vz_v} = \operatorname{MLP}(\operatorname{READOUT}(f^{GNN}(G_{v}; G_{p})), \vw_{t}).
\end{equation}
The parameters of $f^{LM}$ and $f^{GNN}$ are fixed during prompt learning, the parameters of $G_{p}$, the $\vw_{t}$ and the task head (\textit{e.g.}, the MLP layers) are all trainable. 

\subsection{Model Inference}
Finally, we incorporate the pretrained models $f^{LM}$, $f^{GNN}$, and the prompt graph $G_{p}$ in the model inference stage.
We outline three inference methods. The first utilizes the trained LM, referred to as \textbf{P2TAG (LM)}. 
It computes the output of the [CLS] token of the LM for each node in the graph via a mini-batch manner. 
The second method involves further feeding the node [CLS] outputs from P2TAG (LM) into a GNN encoder, referred to as \textbf{P2TAG (GNN)}.
In the training stage, the node representation of $v_i$ is estimated according to the sampled neighborhood in the subgraph, and thus, varies with the sampling process.
The randomness introduced by the sampler is unpleasant during the test stage.
To get the exact node representation, we use the full neighbor sampling strategy when performing inference.
That is, The node representation of $v_i$ at the $\ell$-th layer is obtained by aggregating the representations of its full neighborhood at the $(\ell-1)$-th layer.
The computation process is performed layer by layer.
Taking the node representations at the $(\ell-1)$-th layer as the input, we can compute a batch of node representations at the $\ell$-th by first loading their full neighborhoods and then aggregating the representations of those neighbor nodes at the $(\ell - 1)$-th layer.
The third method involves conducting few-shot node classification through prompting, as described in Section 3.3.
The \textbf{P2TAG} inference loop of our framework is listed in Algorithm~\ref{alg:inference}.

\begin{algorithm}[htbp]
    \caption{The inference of P2TAG}
    \label{alg:inference}
    \textbf{Input} TAG $G=(\mathcal{V}, \mathcal{E}, \mathcal{S})$, batch size $b$, the number of GNN layers $L$\;
    \textbf{Output} Node embeddings\;

    \tcc{Obtain [CLS] embeddings from LMs for all nodes in a mini-batch manner.}
    $\mX \gets \bm{0}, i \gets 0, j \gets \min(b, n)$\;
    \While{$i < j$}{
      $\mO = f^{LM} (\mS [i:j])$\;
      $\mX [i:j] = \mO [:, 0]$\;
      $i \gets j, j \gets \min (j + b, n)$
    }

    \tcc{initialize the graph and text prompt.}
    $\vp_i^{(0)} = 
    \begin{cases}
    f^{LM}(\vs_{y_i}) & \text{if } i \leq N, \\
    \text{Random Initialization} & \text{otherwise}.
    \end{cases}$
    $\vw_t^{(0)} \gets f^{LM} (\vs_{v});$
    
    \tcc{Propagate node embeddings via a GNN encoder layer by layer.}
    $\mH^{(0)} \gets [\mX; G_{p}]$  \\
    \For {$\ell \gets 1$ \KwTo $L$} {
        $\mH^{(\ell)} \gets \bm{0}, i \gets 0, j \gets \min(b, n)$\;
        \While{$i < j$}{
          $\mH^{(\ell)} [i:j] = \relu (\hat{\mA} [i:j] \mH^{(\ell - 1)} \mW^{(\ell)})$\;
          $i \gets j, j \gets \min (j + b, n)$
        }
    }
    $\mH^{(L)} \gets [\mH^{(L)};w_{t}]$ \\
    $\mZ \gets \operatorname{MLP}(\mH^{(L)})$ \\
    \Return{Probabilities of the classes $\mZ$}\;
\end{algorithm}

\hide{
\subsection{Complexity Analysis}
For model training, we first construct a mini-batch graph with batch size $b$ and maximum sequence length $K$ with at most $b^2$ edges between these nodes. 
For the language model, it will cost $O(b \times (K d^2 + K^2 d))$ time to compute the output embeddings. 
The GNN encoder will take at most $O(b \times d^2 + b^2 \times d)$. 
Therefore, the overall training time of each batch is $O(b \times (K d^2 + K^2 d))$ since the computation cost of the GNN encoder is negligible to the language model. 
For model inference, language model costs $O((N/b) \times b \times (K d^2 + K^2 d) = O(NKd^2 + NK^2d)$. 
As for the space complexity, the DeBERTa needs $O(bK^2 + bKd + bkd)$ memory for each batch with size $b$ and maximum relative distance $k$, while the space complexity of GNNs can be negligible. 
}

\section{Experiment}
\label{sec:exp}

\begin{table}[tp]
    \centering
    \caption{Dataset statistics of the used TAGs.}
    \label{tab:dataset_info}
    \setlength{\tabcolsep}{3.2mm}{
    \begin{tabular}{lrrc}
        \toprule
        Dataset & \#Nodes     & \#Edges & Avg. Degree \\
        \midrule
        ogbn-arxiv & 169,343 & 1,166,243 & 13.7 \\
        ogbn-products & 2,449,029 & 61,859,140 & 50.5 \\
        \midrule
        Children & 76,875 & 1,554,578 & 40.4 \\
        History & 41,551 & 358,574 & 17.2 \\
        Computers & 87,229 & 721,081 & 16.5 \\
        Photo & 48,362 & 500,928 & 20.7  \\
        \bottomrule
    \end{tabular}
    }
    \normalsize
    \vspace*{-10pt}
\end{table}

\begin{table*}[ht]
\centering
\captionsetup{skip=6pt}
\caption{The accuracy of few-shot node classification on two OGB datasets. }
\label{tab:fw_ogb}
\setlength{\tabcolsep}{0.9mm}{
\begin{tabular}{l|cc|cc|cc|cc} 
\toprule
            Dataset    & \multicolumn{4}{c|}{ogbn-arxiv} & \multicolumn{4}{c}{ogbn-products} \\
\cline{1-9}
            Setting    & 5-way 3-shot & 5-way 5-shot & 10-way 3-shot & 10-way 5-shot & 5-way 3-shot & 5-way 5-shot & 10-way 3-shot & 10-way 5-shot \\
\midrule
OGB features  & 53.24±0.77 & 59.31±0.71 & 38.62±0.85 & 44.59±0.61 & 44.15±1.30 & 50.57±1.83 & 32.79±0.88 & 38.75±0.61\\
GPN    & 61.65±0.67 & 66.34±0.78 & 47.36±0.25 & 53.60±0.86 & 64.04±1.92 & 69.86±1.90 & 52.68±0.66 & 58.36±0.45\\
G-Meta & 60.06±1.56  & 63.77±2.53 & 48.36±0.94 & 52.78±1.04 & 64.06±1.18 & 66.02±0.71 & 51.17±1.48 & 57.81±1.40\\
TENT   & 62.24±0.35  & 65.80±0.37 & 45.08±0.54 & 51.84±0.86 & 62.69±2.78 & 66.63±1.48 & 47.63±0.59	& 51.53±0.90\\
\midrule
GIANT  & 73.50±1.59 & 78.18±1.25 & 60.05±1.20 & 64.96±0.69 & 69.44±1.40 & 75.46±1.53 & 59.23±0.64 & 65.67±0.38\\
G2P2   & 70.93±1.29  & 73.07±1.22 & 56.73±0.68 & 59.47±0.69 & > 30 days & > 30 days & > 30 days & > 30 days \\
\midrule
\model (LM)  & \uline{78.14±1.18} & \uline{81.33±1.34} & 65.21±1.26 & 69.90±0.94 & 75.38±1.27 & 78.66±1.55 & 61.87±0.91 & 67.06±0.49\\
\model (GNN) & 77.73±1.76 & 81.31±0.84 & \uline{66.21±0.82} & \uline{70.42±0.62} & \textbf{80.95±1.61} & \uline{83.65±1.24} & \uline{70.04±0.33} & \uline{73.67±0.35} \\
\model  &  \textbf{80.18±0.27} & \textbf{84.76±0.89} & \textbf{67.81±0.97} & \textbf{70.88±0.96} & \uline{80.54±1.59}	& \textbf{83.77±1.45} & \textbf{71.14±0.26} & \textbf{75.99±0.41}\\ 
\bottomrule
\end{tabular}
}
\end{table*}

In this section, 
we conduct comprehensive experiments to validate the effectiveness of our framework, particularly emphasizing the enhanced performance for few-shot node classification tasks.

\subsection{Experimental Setup}

\vpara{Datasets.}
We conduct experiments across two categories of datasets. All these datasets are publicly available and widely used in TAG-related research. 
The statistics of the datasets are given in Table~\ref{tab:dataset_info}. 
\begin{itemize} [leftmargin=15pt]
    \item \textbf{Open Graph Benchmark (OGB)}. \textit{ogbn-arxiv} is a directed citation graph between all computer science (CS) ArXiv papers indexed by Microsoft Academic Graph (MAG)~\cite{wang2020microsoft}. 
    We convert the abbreviated labels into their full forms as specified on the arXiv website, such as transforming "NA" into "Numerical Analysis".
    \textit{ogbn-products} is an undirected and unweighted graph representing an Amazon product co-purchasing network~\cite{Bhatia16}. 
    \item \textbf{Amazon Review}. Amazon Review includes graphs composed of products, which are constructed based on co-purchased and co-viewed patterns. 
    We use \textit{Children}, \textit{History}, \textit{Computers} and \textit{Photo} datasets compiled by~\cite{yan2023comprehensive}.
    For the four datasets, we extract bag-of-words features and utilize principal component analysis (PCA) to reduce dimensions, generating a 100-dimensional feature for each node.
\end{itemize}
In terms of the class split, ogbn-arxiv adopts the same partitioning strategy as TENT~\cite{wang2022task}. Meanwhile, for other datasets, we employ a random division approach. Notice that our \model differs from meta-learning methods, as it does not require the establishment of specific training and validation tasks. Instead, we ensure fairness using the same test tasks as other baselines.

\begin{table*}[!ht]
\centering
\captionsetup{skip=6pt}
\caption{The accuracy of few-shot node classification on four Amazon Review datasets.}
\begin{minipage}{\textwidth}
\centering
\setlength{\tabcolsep}{3.8mm}{
\begin{tabular}{l|ccc|ccc} 
\toprule
            Dataset & \multicolumn{3}{c|}{Childern} & \multicolumn{3}{c}{History} \\
\cline{1-7}
            Setting & 3-way 3-shot & 3-way 5-shot & 3-way 10-shot & 3-way 3-shot & 3-way 5-shot & 3-way 10-shot \\
\midrule
node features & 39.68±0.61 & 43.33±1.78 & 49.00±0.77 & 37.05±1.63 & 38.16±1.33 & 40.82±1.07 \\
GPN  & 54.77±0.42 & 60.51±1.16 & 63.89±0.93 & 38.48±0.74 & 40.63±0.82 & 43.88±1.01 \\
G-Meta & 54.00±2.38 & 57.76±1.43 & 61.62±1.52 & 40.41±1.32 & 41.11±0.76 & 42.50±0.86 \\
TENT   & 53.23±0.76 & 60.52±1.73 & 64.32±0.81 & 38.30±0.72 & 37.73±0.43 & 41.47±2.05 \\
G2P2   & 51.33±1.84 & 55.88±1.18 & 58.85±0.84 & 47.21±0.37 & 49.89±0.97 & 55.63±1.11\\
\midrule
\model (LM)  & \uline{73.64±1.13} & \uline{77.43±1.46} & 80.53±1.27 & \uline{68.88±2.10}	& \uline{72.36±1.28} & \uline{76.81±0.47}  \\
\model (GNN) & 73.20±1.34 & 76.71±0.82 & \uline{80.67±1.54} & 63.33±1.00 & 68.15±1.41 & 72.68±0.98 \\
\model  &  \textbf{77.35±1.09} & \textbf{81.43±1.24} & \textbf{84.67±1.90} & \textbf{72.35±1.85} & \textbf{76.19±1.25} & \textbf{80.96±0.74}  \\
\bottomrule
\end{tabular} 
}
\end{minipage}

\vspace{1pt}

\begin{minipage}{\textwidth}
\centering
\setlength{\tabcolsep}{3.8mm}{
\begin{tabular}{l|ccc|ccc} 
\toprule
            Dataset   & \multicolumn{3}{c|}{Computers} & \multicolumn{3}{c}{Photo} \\
\cline{1-7}
            Setting    & 3-way 3-shot & 3-way 5-shot & 3-way 10-shot & 3-way 3-shot & 3-way 5-shot & 3-way 10-shot \\
\midrule
node features & 37.19±1.36 & 40.11±1.72 & 44.55±1.11 & 40.43±0.64 & 44.41±1.68 & 50.57±1.38\\
GPN  & 71.73±1.67 & 70.47±1.72 & 71.56±1.91 & 74.73±1.75 & 76.01±1.12 & 70.77±0.97\\
G-Meta & 71.58±2.19 & 71.14±2.52 & 72.36±0.63 & 69.68±1.08 & 72.41±1.09 & 73.54±1.22\\
TENT   & 60.68±1.94	& 62.13±1.17 & 65.96±1.92 & 68.24±1.20 & 71.20±1.30 & 72.19±2.03\\
G2P2   &  61.36±1.20 & 65.13±1.90 & 68.56±1.26 & 70.97±1.51 & 72.49±1.23 & 76.52±2.37\\
\midrule
\model (LM)  & 66.73±2.16 & 71.17±0.54 & 74.77±1.06 & 73.49±1.77 & 75.68±0.99 & 77.89±0.94  \\
\model (GNN) & \uline{86.95±1.71} & \uline{88.67±0.51} & \uline{92.15±0.88} & \uline{86.47±1.45} & \uline{88.20±0.49} & \textbf{90.72±0.58}  \\
\model  & \textbf{87.24±1.38} & \textbf{89.55±0.54} & \textbf{93.76±0.63} & \textbf{86.73±2.14} & \textbf{88.73±0.98}	& \uline{90.68±0.76}  \\
\bottomrule
\end{tabular} 
}
\end{minipage}
\label{tab:fw_amazon}
\vspace{-5pt}
\end{table*}

\vpara{Compared methods.}
We choose three types of methods for comparison, all of which can address the problem of few-shot node classification. 
These include methods based on meta-learning on graphs, self-supervised pre-training methods on text attribute graphs, and methods following the paradigm of pre-training and prompting. 
Specifically, meta-learning methods such as GPN~\cite{ding2020graph}, G-Meta~\cite{huang2020graph} and TENT~\cite{wang2022task}; GIANT~\cite{ChienCHYZMD22}, proposed \model (LM) and \model (GNN) as self-supervised pre-training methods; G2P2 \cite{wen2023augmenting} follows the same pre-training and prompting paradigm as proposed \model. 
In the pre-training phase, we concurrently train both the LM and GNN, considering two scenarios during inference. 
\model (LM) derives vector representations from the original text of nodes. Meanwhile, \model (GNN) leverages these node representations as inputs to the GNN, resulting in the output obtained.

\vpara{Evaluation.}
GPN, G-Meta, and TENT are methods based on meta-learning, where they are learned on train tasks and are subsequently evaluated on test tasks. 
In contrast, the proposed \model along with G2P2 and GIANT, are evaluated exclusively on test tasks. Specifically, \model (LM), \model (GNN), and GAINT infer the classes of nodes in the query set by fitting a logistic regression classifier on the support set. 
Furthermore, \model and G2P2 extend their approach by constructing prompts through the support set.
For a detailed comparison, we conduct experiments in four different settings for OGB datasets: 5-way 3-shot, 5-way 5-shot, 10-way 3-shot, and 10-way 5-shot. 
Considering the number of labels, we use three settings for Amazon Review datasets: 3-way 3-shot, 3-way 5-shot, and 3-way 10-shot.

\vpara{Parameter configuration.}
For our framework, the selection of LMs and GNNs is flexible. 
In our experiment, we choose a representative LM --- DeBERTa-base and a powerful GAT model for the main experiments. 
The DeBERTa-base is a 100M-parameter pre-trained language model with a hidden size of 768. 
We keep the same hidden size of the GAT model with DeBERTa-base. 
We also explore other LMs in the ablation studies. 
We use AdamW optimizer~\cite{loshchilov2017decoupled} with learning rate lr = 1e-5 for model optimization. 
We run 3 epochs for all datasets.
We construct 5 groups of test tasks for each $N$-way $K$-shot setting, with each group consisting of 50 tasks specifically formulated from the test set label. 
In each task, the support set length is $K$, and the query set length $Q$ is set to 10.
We calculate the average performance within each group and report the overall mean and standard deviation across these 5 groups of tasks.

\subsection{Performance Analysis}

Our main results are summarized in Table~\ref{tab:fw_ogb} and Table~\ref{tab:fw_amazon}. 
The proposed \model (LM) outperforms the meta-learning methods, increasing the average accuracy on six datasets from +2.27\% $\sim$ +35.51\%; 
the \model (GNN) achieves an average improvement with +16.38\% $\sim$ +27.55\%; 
the \model performs best on most datasets, with an average improvement of +18.98\% $\sim$ +35.98\%.
Compared with the pre-training method that also utilize raw text such as GIANT and G2P2, our \model still has better performance, which demonstrates its effectiveness. 
On the ogbn-arxiv dataset, although G2P2 constructs prompts on the basis of pre-training, its performance is lower than that of GIANT, \model (LM) and \model (GNN). 
This underscores the importance of a well pre-trained model for few-shot node classification. 
Pre-training on TAGs often comes with increased time expenditure, such as G2P2 taking over 30 days on the ogbn-products dataset. 
The proposed \model employs a more general approach to jointly train LM and GNN, consuming less than one day on this dataset.
The \model, employing mixed prompts, further enhances performance on the pre-trained model \model (LM) and \model (GNN). 
It secures the best outcomes on most datasets, with an average enhancement of 2.89\% compared to \model (GNN), demonstrating the effectiveness of prompts.
On the History dataset, \model (LM) achieves the second-best results; however, after passing through the GNN encoder, \model (GNN) experiences an average of 4.63\% decrease. 
This might be attributed to the quality of the topological structure within the data. 
This hypothesis is further supported by the minimal improvement observed when comparing meta-learning methods that do not utilize raw texts (GPN, G-Meta, TENT) to the node features method. 
The proposed graph prompt improvement strategy, which initializes through label text, mitigates this issue. 
On the History dataset, \model further enhances performance compared to \model (LM), achieving an average improvement of 3.8\% across three settings.

\begin{table}
\centering
\captionsetup{skip=6pt}
\caption{Ablation study of language models on ogbn-arxiv dataset. \textmd{We choose various LMs, then report the classification accuracy achieved with \model (LM).}}
\label{tab:ablation_lm}
\setlength{\tabcolsep}{0.65mm}{
\begin{tabular}{l|cc|cc} 
\toprule
        \multirow{2}*{Setting}    & \multicolumn{2}{c|}{5-way} & \multicolumn{2}{c}{10-way} \\
\cline{2-5}
            & 3-shot & 5-shot & 3-shot & 5-shot  \\
\midrule
DeBERTa-base  & 78.14±1.18 & 81.33±1.34 & 65.21±1.26 & 69.90±0.94  \\
DeBERTa-large & \textbf{78.64±1.24} & \textbf{83.55±1.05} & \textbf{66.00±1.12} & \textbf{71.96±1.06} \\
e5-v2-base &  78.62±1.00 & 82.44±1.22 & 65.79±0.64 & 71.02±0.60   \\
e5-v2-large   & 77.31±1.68 & 81.87±1.09 & 64.60±0.72 & 70.08±0.60  \\
\bottomrule
\end{tabular}
}
\end{table}

\subsection{Ablation Studies}
In the previous sections, we demonstrate the powerful performance of the \model (LM), \model (GNN), and \model. 
This part analyzes the impact of the different types of prompting and LMs.

\vpara{Effect of LMs.}
To better analyze the impact of LMs, we explore other LMs such as e5-v2-base with 110M parameters~\cite{wang2022text}.
We also try larger LMs such as DeBERTa-large with 350M parameters and e5-v2-large with 330M parameters. 
The results are reported in Table~\ref{tab:ablation_lm}. 
Generally, the results of LMs are quite similar, with differences within 1.5\%.
The reason that e5-large does not achieve better results may be attributed to insufficient training iterations.
This paper selects DeBERTa-base, intending to address the joint learning problem of LMs and GNNs in a more general manner. 
There remains room for further exploration in the specific choice of LMs.

\begin{figure}[!htbp]
    \captionsetup{skip=6pt}
    \centering
    \includegraphics[width=0.45\textwidth]{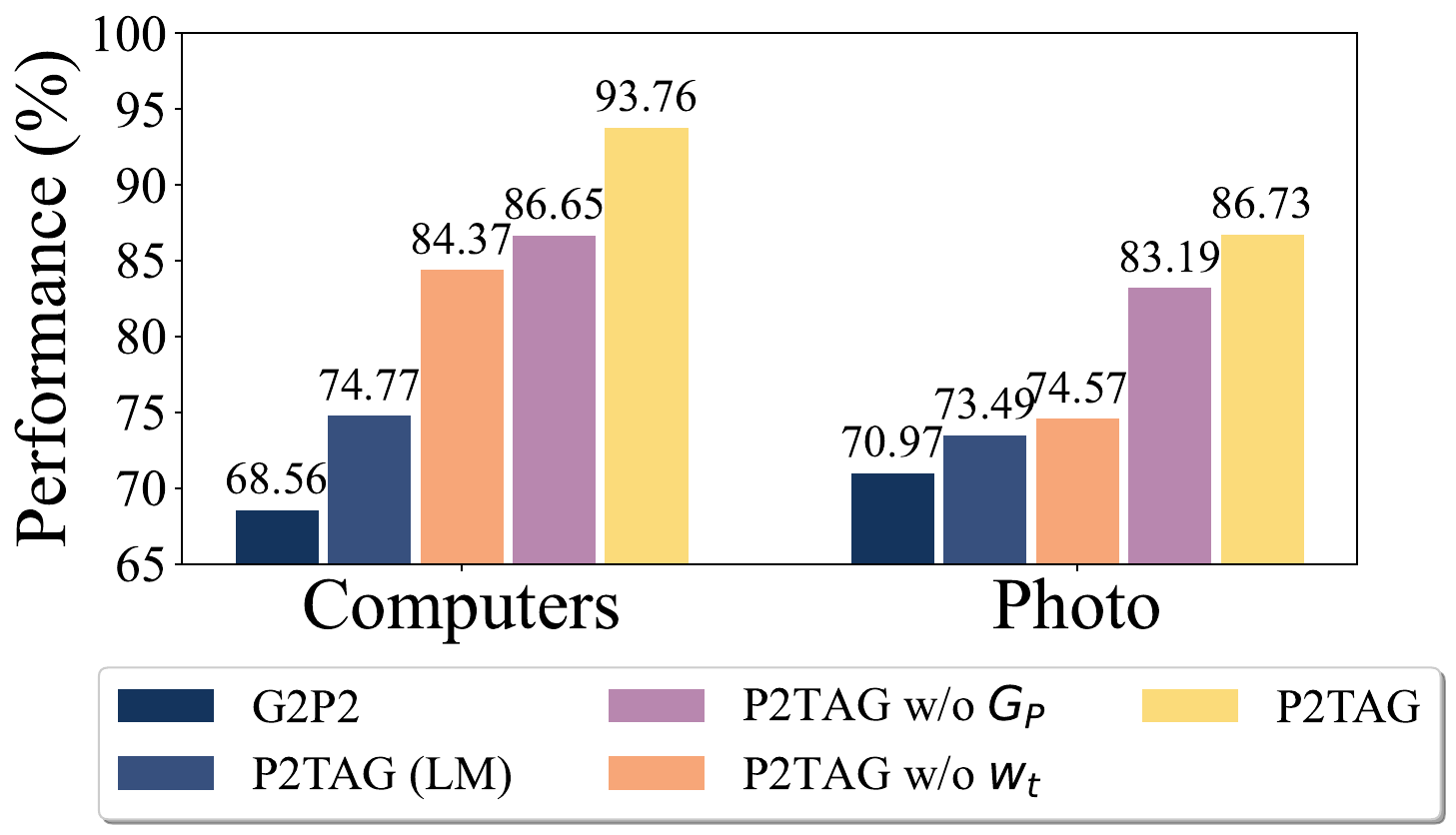}
    \caption{Effect of different prompt types. \textmd{We omit label initialization on the prompt graph (w/o $P_{G}$) and node text embedding (w/o $\mathbf{w}_{t}$) and report the classification results on two datasets.}}\label{fig:ab_prompt_type}
\end{figure}

\vpara{Effect of different prompt types.}\label{sec:effct_of_prompts}
In Section~\ref{sec:graph_prompt}, we discussed two significant enhancements. Here, we further investigate these improvements by dissecting them and presenting the findings in Figure~\ref{fig:ab_prompt_type}. We use the \model w/o ${P_{G}}$ when the label text embedding is not utilized to initialize our prompt graph, and the \model w/o $\mathbf{w}_{t}$ when node text embedding is not employed to bolster the prompted graph embedding (\textit{i.e.}, the graph embedding is directly inputted into MLP layers). We report the results on Computers and Photo datasets. 
The results reveal that our prompting approach outperforms both the \model (LM) and baseline G2P2. 
Furthermore, the absence of ${P_{G}}$ has a marginally greater impact than omitting ${\mathbf{w}_{t}}$ across both datasets, underscoring the importance of structural effectiveness in the TAG context.

\begin{table}[t]
\centering
\captionsetup{skip=6pt}
\caption{The accuracy of classification by \model with different hyperparameters on ogbn-arxiv dataset. \textmd{We report the results with \model (GNN) and \model, where the underscored annotations indicate the parameters used in the main result tables.}}
\label{tab:ablation_hp}
\begin{tabular}{lrrr} 
\toprule
 \multicolumn{2}{l}{Setting}   & 5-way 5-shot & 10-way 5-shot  \\
\midrule
\multirow{4}*{\parbox{2.7cm}{\model (GNN):\\ sequence length}} & 32  &  79.74±1.32 & 67.81±0.48    \\
& 64  &   81.40±1.26 & 69.61±0.52          \\
& \uline{128} &  81.31±0.84	& \textbf{70.42±0.62}        \\
& 256 &  \textbf{81.61±0.95}	& 70.40±0.70             \\
\midrule
\multirow{4}*{\parbox{2.7cm}{\model (GNN):\\ mask rate}} & 0.15 &  43.93±0.41 & 27.02±0.48             \\
& 0.30 &   60.66±1.00 & 46.34±0.48          \\
& 0.50 &   80.48±1.10 & 69.96±0.69  \\
& \uline{0.75} &   \textbf{81.31±0.84} & \textbf{70.42±0.62}      \\
\midrule
\multirow{4}*{\parbox{2.7cm}{\model (GNN):\\ walk length}} & 5 & 81.28±1.07	& 70.18±0.54             \\
& \uline{10} &   81.31±0.84	& \textbf{70.42±0.62}        \\
& 20 &  81.28±1.43 & 70.20±0.59             \\
& 50 & \textbf{81.70±0.61} & 70.11±0.74 \\
\midrule
\multirow{4}*{\parbox{2.7cm}{\model :\\ token num}} & 2 & 84.31±0.95 & 70.04±1.42 \\
& \uline{5} & \textbf{84.76±0.89} & 70.16±0.78 \\
& \uline{10} & 84.35±1.07 & \textbf{70.88±0.96} \\
& 20 & 84.06±1.12 & 69.71±0.69 \\
\bottomrule
\end{tabular}
\vspace*{-5pt}
\end{table}

\subsection{Hyperparameter Analysis}
In this part, we conduct a comprehensive analysis of hyperparameters to elucidate their impact on the performance of our framework. We analyze the four key hyperparameters in our experiments. 
The "sequence length" means the reserved length of raw texts. The "mask rate" represents the proportion of masked tokens in training. The "walk length" means the length of random walks used in the GraphSAINT sampler for mini-batch training. 
The "token num" denotes the number of tokens in the graph prompt. 
The results of hyperparameter experiments are shown in Table~\ref{tab:ablation_hp}. 

\vpara{Preprocessing: length of truncated sequences.}
Since each batch will be fed to both the language model and the GNN encoder in training, we need to increase the batch size to better leverage the ability of GNNs. 
However, the batch size is limited due to the large memory cost of the LM. 
Therefore, considering both training efficiency and information loss, we explore using a smaller truncated length in our experiments to utilize a larger batch size. 

\vpara{Pre-training: mask rate.}
A smaller mask rate represents an easier self-supervised task and might make the model overfitted. 
To make the self-supervised task harder, we need to choose a large enough mask rate to let the model learn better. 
From the results, we find that our framework achieves the best performance with a 75\% mask rate on the ogbn-arxiv dataset.

\vpara{Sampling: lengths of random walks.}
For each batch, the sampler conducts multiple random walks from different root nodes with the same length. 
The GNN encoder of our framework relies on the topology of the sampled mini-batch graph to propagate and aggregate the information between nodes. 
Therefore, the length of random walks used to construct the mini-batch will influence the model performance. 

\vpara{Prompting: number of tokens.} Incorporating the prompt graph as a trainable parameter highlights the importance of token quantity for downstream tasks. Intuitively, more tokens might create a richer internal structure, offering enhanced information. However, managing more parameters often complicates convergence in few-shot scenarios. To accurately incorporate the textual information of labels into prompting, we typically set the number of tokens to match the number of classes in the task. 
Surprisingly, 2 tokens also yield results, implying a minimal internal structure. This suggests the effectiveness of token features in improving performance.

\section{Related Work}
\label{sec:related}
In this section, we introduce the related work, including graph representation learning and few-shot node classification. 

\subsection{Graph Representation Learning}
Graph neural networks (GNNs) provide the foundation for applying deep learning on graphs and yielding good results on several downstream tasks. 
The earlier works~\cite{welling2016semi, VelickovicCCRLB18, xu2019graph, defferrard2016convolutional} perform convolution on small-scale graphs using all topological relations in a semi-supervised way. 
The subsequent works focus on sampling strategies \cite{HamiltonYL17, chiang2019cluster, ZengZSKP20} and model architecture \cite{wu2019simplifying, rossi2020sign} to enhance the scalability of GNNs and apply them to large-scale graphs. 
GraphSAGE~\cite{HamiltonYL17} first proposed the idea of neighborhood sampling,
and later it was applied in a real-world recommendation system by PinSAGE~\cite{ying2018graph}. 
GraphSAINT~\cite{ZengZSKP20}, 
first samples subgraphs~\cite{leskovec2006sampling} and runs full-batch GNNs on sampled subgraphs. 
Additionally, there are works ~\cite{zhao2022hierarchical,zhao2023hierarchical} focusing on the hierarchy within the graph to obtain better representations.
The self-supervised learning methods on GNNs are developed via contrastive and generative ways. 
The contrastive methods~\cite{zeng2021contrastive, you2020graph, thakoor2022large, hassani2020contrastive, qiu2020gcc, zhu2020deep} adopt data augmentation with or without negative sampling to construct samples to optimize the contrastive loss under different augmentations. 
GRACE~\cite{zhu2020deep} aims to maintain node uniformity across views. 
BGRL~\cite{thakoor2022large} designs two encoders for two views with data augmentation. 
Some studies focus on graph generative learning.
GraphMAE~\cite{HouLCDYW022} and GraphMAE2~\cite{hou2023graphmae2} are proposed to reconstruct masked attributes of corrupted nodes for representation learning.

The semantics and graph topology of TAGs can express real-world relationships between entities or objects. 
Most GNNs do not consider the text processing in the TAG but directly use the numerical features, which are generated through text encoding, as attributes of nodes. 
Recent studies~\cite{ChienCHYZMD22, YasunagaLL22} utilize graph topology to enhance the representation during text pre-training. Despite the promising results, the language model is still independent of the GNNs~\cite{zhu2021textgnn, li2021adsgnn}. 
GraphFormer~\cite{yang2021graphformers} designs a nested architecture of GNNs and Transformers. GLEM~\cite{zhao2022learning} implements the fusion of LMs and GNNs on large-scale text-attributed graphs with the variational expectation-maximization framework.
Although these methods progress in integrating LMs and GNNs, there is still a lack of discussions on self-supervised learning on large-scale text-attribute graphs. 
Our proposed pre-train framework enhances LMs utilizing GNNs and achieves joint training with the objective of self-supervision. 
\citet{yan2023comprehensive} release the benchmark on TAGs, and provide multiple clean TAG datasets. These datasets also lay a solid foundation for our experiments.

\subsection{Few-shot Node Classification}
Few-shot node classification on graphs to categorize nodes within a graph with limited labeled nodes. 
Drawing inspiration from the success of meta-learning in few-shot classification tasks within computer vision~\cite{finn2017model, SnellSZ17}, several studies apply meta-learning techniques to graph-based tasks. 
GFL~\cite{yao2020graph} and GPN~\cite{ding2020graph} utilize prototype networks to learn the distance from nodes to classes. 
Meta-GNN~\cite{zhou2019meta} optimizes the graph neural network with model-agnostic meta-learning. 
G-Meta~\cite{huang2020graph} addresses the meta-learning problem on both single and multiple graphs by extracting local subgraphs. 
TENT~\cite{wang2022task} enhances the model's generalization capabilities by adapting at three levels: nodes, edges, and tasks. 

With the increasing attention towards LLMs, several works attempt to follow the paradigm of pre-training and prompt tuning to address the problem of few-shot node classification. 
Prog~\cite{sun2023all} incorporates both node-level and edge-level tasks by constructing prompts at the graph level, but lacks analysis of the text attribute. 
ENG~\cite{yu2023empower} employs LLM with designed prompts to refine node texts and reconstructs the adjacency matrix semantically, which is then used as input for the GNN. 
G2P2~\cite{wen2023augmenting} enhances text information through graph structure, aligning text representations in three forms during the pre-training phase. In the tuning phase, it uses the neighborhood text of the target node and label text to generate the initial parameters of prompts while freezing the parameters of the LM and GNN during the tuning process.

\section{Conclusion}
\label{sec:conclusion}

Our paper focuses on few-shot node classification on the TAG. 
We address this problem by employing a graph pre-training and prompting approach. 
The proposed framework \model utilizes a masked language modeling objective for the joint training of the language model and GNN model. 
We also propose a new prompting method that mixes graph and text information, enabling the pre-trained model on TAG to better adapt to downstream few-shot node classification tasks.
We conduct experiments on six real-world TAG datasets and our \model framework achieves state-of-the-art results on the six datasets with +18.98\% $\sim$ +35.98\% improvement.

\section{Acknowledgement}
This work is supported by National Key R\&D Program of China 2021ZD0113304, Natural Science Foundation of China (NSFC) 62276 \ 148 and 62425601, CCF-Zhipu AI Large Model Fund (Grant 202213), Zhipu AI - Anhui University Joint Research Center on Foundation Model and the University Synergy Innovation Program of Anhui Province (GXXT-2023-050), the New Cornerstone Science Foundation through the XPLORER PRIZE, and Tsinghua-Bosch Joint ML Center.

\clearpage

\balance

\bibliographystyle{ACM-Reference-Format}
\bibliography{reference}

\clearpage

\appendix

\section{Appendix}

\subsection{Implementation Notes}

\vpara{Running environment.}
The experiments are conducted on a Linux machine with AMD EPYC 7642 48-Core Processor, 1T RAM, and 8 NVIDIA A100 (80G). 
Our code is implemented with PyTorch 1.10 and Python 3.8.

\vpara{Model Configuration.}
Our framework involves the joint training of LM and GNN, with some of the key parameters presented in Table~\ref{tab:model_conf}. 
For more detailed configurations, please find in our code. 

\begin{table}[!hp]
\centering
\captionsetup{skip=6pt}
\caption{Key parameters of the \model.}
\label{tab:model_conf}
\begin{tabular}{ll} 
\toprule
 Parameters  & Value        \\
\midrule
language model  &  microsoft/deberta-base     \\
sequence length &  128   \\
mask rate       &  0.75  \\
\midrule
GNN encoder       &  graph attention network \\
GNN hidden size &  768   \\
walk length     &  10  \\
number of root  &  10  \\
\midrule
length of query set & 10 \\
number of test groups & 5 \\
number of task in each group & 50 \\
\midrule
number of tokens & the number of classes of task \\
prompt epochs & 50 \\
\bottomrule
\end{tabular}
\end{table}

\subsection{Comparing \model with GPrompt}
Though P2TAG shares a similar task and intuition with GPrompt~\cite{xuanwen2023prompt}, there are significant differences in the prompt design. Since the prompting parts are crucial for the pre-training and prompting paradigm, solely considering the task and intuition may not be fair. Therefore, we further elaborate our design here:

\textbf{Difference in Fundamental Settings of Prompting.} GPrompt uses the entire GNN model as the prompt, which requires fine-tuning the GNN for each downstream task. In contrast, we freeze the GNN parameters and instead use an auxiliary prompt graph to alter the input. This approach, we believe, introduces a significant change to the community.
Our method not only preserves the knowledge encapsulated within the GNN model but also reduces the number of training parameters to a minimal set of node embeddings, thereby enhancing computational efficiency.

\textbf{Novel Label Text Prompt Initialization Strategy.}
Furthermore, we introduce a novel strategy for initializing the prompt graph by encoding the label texts to embeddings, which is a unique and effective element in the TAG scenario. By incorporating the label text embedding, we can easily initialize the prompt graph and align the label space more closely with the downstream task.
For example, a straightforward text prompt design might be: ``Given labels \textit{student}, \textit{teacher}, \textit{worker}, the person with attributes [feature] belongs to [cls]''. In the graph domain, we initialize the prompt graph with these label text embeddings. As a result, the target node embedding naturally aligns closely with the label text embedding via message passing, replicating the prompt initialization process. Moreover, the label text embeddings within the prompt graph can adaptively influence the target node, potentially offering more effectiveness than static text prompts with uniform contributions.

\subsection{Baselines}
We conduct five baselines, among which GPN, G-Meta, and TENT are based on meta-learning and do not utilize raw texts. GIANT employs raw texts for self-supervised learning on graphs. G2P2 also leverages raw texts and follows the paradigm of pre-training and prompting. All of them address the few-shot node classification problem. For each $N$-way $K$-shot few-shot setting, we sample 5 groups, each containing 50 tasks for testing. To ensure a fair comparison, we save the constructed tasks, and all baselines, including our proposed model, are tested on these tasks. G-Meta is an exception due to its more complex sampling process, but we ensure the same number of testing tasks.

\begin{itemize} [leftmargin=15pt]
    \item GPN\footnote{\url{https://github.com/kaize0409/GPN\_Graph-Few-shot}} applies the prototype network from meta-learning to graphs. We set the number of episodes for training to 2000. 
    \item G-Meta\footnote{\url{https://github.com/mims-harvard/G-Meta}} implements the meta-learning method for graphs at three levels using local subgraphs. To match our number of test tasks, the batch size was set to 50 during the generation of test tasks, and this process is repeated 5 times.
    \item TENT\footnote{\url{https://github.com/SongW-SW/TENT}} tunes the model across multiple tasks. We modify this method by Deep Graph Library (DGL) to enable its application on larger-scale graphs, such as the ogbn-products.
    \item GIANT\footnote{\url{https://github.com/amzn/pecos/tree/mainline/examples/giant-xrt}} performs extreme multi-label classification on the raw texts, resulting in stronger node features. We directly download their pre-trained node features for evaluation on the ogbn-arxiv and ogbn-products datasets.
    \item G2P2\footnote{\url{https://github.com/WenZhihao666/G2P2}} enhances text representation with graph structure, aligning at three levels, and subsequently performs few-shot classification through prompting. We set the hidden layer dimension to 128 and the number of epochs to 3.
\end{itemize}

\subsection{Analysis of \model Inference Cost}

We analyze the inference cost of P2TAG and baselines with respect to the text sequence length $K$, embedding dimension $d$, number of nodes $N$, number of nodes $b$ in the ego-graph, and the number of graph tokens $t$.

The cost of P2TAG (LM) stems from the language model (LM), with a complexity of $O(N \times (Kd^2 + K^2 d))$. 
P2TAG (GNN) additionally utilizes a GNN encoder with negligible computation overhead. The complexity is $O(N \times (Kd^2 + K^2 d))$. 
P2TAG includes two stages: pre-training and prompting. The complexity is $O((N+b+t) \times (Kd^2 + K^2 d))$. 
The complexity of GIANT is $O(N \times (Kd^2 + K^2 d) + rd \log(N))$, where $r$ is a hyperparameter. 
The complexity of G2P2 is $O((N+c+1) \times (Kd^2 + K^2 d))$, where $c$ represents the number of neighbors.

Overall, the proposed P2TAG exhibits computational costs similar to other baselines, yet achieves improved performance in few-shot classification.

\subsection{Ablation Study about the GNN Backbone}

We use GAT due to its ability to adaptively transfer information from our prompt graph to the target node, matching our graph prompt design. Our ablation study on GCN and GraphSAGE within P2TAG on the ogbn-arxiv dataset shows:

\begin{table}[H]
\centering
\captionsetup{skip=6pt}
\caption{Ablation study results on the ogbn-arxiv dataset.}
\label{tab:ablation_backbone}
\resizebox{\linewidth}{!}{%
\begin{tabular}{lcccc}
\toprule
GNN Encoder & 5-way 3-shot (\%) & 5-way 5-shot (\%) & 10-way 3-shot (\%) & 10-way 5-shot (\%) \\
\midrule
GCN         & 75.41 $\pm$ 1.21   & 78.90 $\pm$ 0.61  & 60.40 $\pm$ 1.26   & 64.54 $\pm$ 0.69  \\
GraphSAGE   & 74.02 $\pm$ 1.43   & 77.58 $\pm$ 0.73  & 59.60 $\pm$ 1.40   & 64.45 $\pm$ 0.95  \\
GAT (Ours)  & 80.18 $\pm$ 0.27   & 84.76 $\pm$ 0.89  & 67.81 $\pm$ 0.97   & 70.88 $\pm$ 0.96  \\
\bottomrule
\end{tabular}%
}
\end{table}

GAT outperforms the other encoders, as expected. The performance gap can be attributed to the lack of adaptive selection when introducing information (including noise) from the prompt graph into the target node embedding.

\subsection{Comparing More Classes Settings in Few-Shot Experiments}
We chose the 3-way, 5-way, and 10-way settings across all datasets for two reasons. First, most works on few-shot node classification~\cite{yao2020graph, ding2020graph, huang2020graph, wang2022task} retain these settings, so we chose to follow them. Second, the limited number of classes in the datasets, such as the Amazon dataset with only slightly more than 10 classes, restricts us to these settings. Some meta-learning methods require splitting the classes into training, validation, and test sets, which limits us to the 3-way setting.
We further conducted new experiments on the ogbn-arxiv dataset under the 15-way setting for nine models (including variants). The results are as follows:

\begin{table}[H]
\centering
\small
\caption{Comparison under the 15-way setting on the ogbn-arxiv dataset.}
\resizebox{0.8\linewidth}{!}{%
\begin{tabular}{lcc}
\toprule
Model          & 15-way 3-shot (\%) & 15-way 5-shot (\%) \\
\midrule
Raw Features   & 32.24 $\pm$ 0.26    & 37.37 $\pm$ 0.82   \\
GPN            & 38.38 $\pm$ 0.39    & 44.46 $\pm$ 0.68   \\
G-Meta         & 39.36 $\pm$ 0.28    & 44.85 $\pm$ 0.48   \\
TENT           & 37.61 $\pm$ 0.32    & 40.20 $\pm$ 0.37   \\
GIANT          & 52.77 $\pm$ 0.50    & 57.74 $\pm$ 0.86   \\
G2P2           & 48.48 $\pm$ 0.97    & 52.24 $\pm$ 0.85   \\
P2TAG (LM)     & 57.31 $\pm$ 0.57    & 62.71 $\pm$ 0.66   \\
P2TAG (GNN)    & 58.49 $\pm$ 0.53    & 63.60 $\pm$ 0.68   \\
P2TAG          & 59.42 $\pm$ 0.42 (6.65 $\uparrow$) & 64.05 $\pm$ 0.56 (6.31 $\uparrow$) \\
\bottomrule
\end{tabular}%
}
\end{table}

We can find that increasing the number of classes indeed makes the few-shot node classification task more challenging for all models. For example, the raw feature method achieves 59.31\% in the 5-way setting, but only 44.59\% in the 10-way setting. Nevertheless, P2TAG consistently outperforms the baselines, improving over the next best method by 6.65\% and 6.31\%, respectively.

\end{document}